\documentclass[aps,prx,reprint,groupedaddress]{revtex4-1} 
\usepackage{graphicx}
\usepackage{amsmath,amssymb}
\usepackage{textcomp,siunitx}
\usepackage{ulem}
\DeclareSIUnit\gauss{G}
\usepackage[
	pdftex,
	colorlinks=true,
  	urlcolor=blue,
  	linkcolor=blue,
  	citecolor=blue
]{hyperref}
\usepackage{color}

\newcommand{\avg}[1]{\left\langle #1 \right\rangle}%
\renewcommand{\vec}[1]{\boldsymbol{#1}}

\begin{document}

\title{Subdiffusion and heat transport in a tilted 2D Fermi-Hubbard system}
\author{Elmer Guardado-Sanchez}
\altaffiliation{These authors contributed equally to this work.}
\affiliation{Department of Physics, Princeton University, Princeton, New Jersey 08544 USA}
\author{Alan Morningstar}
\altaffiliation{These authors contributed equally to this work.}
\affiliation{Department of Physics, Princeton University, Princeton, New Jersey  08544 USA}
\author{Benjamin M. Spar}
\affiliation{Department of Physics, Princeton University, Princeton, New Jersey  08544 USA}
\author{Peter T. Brown}
\affiliation{Department of Physics, Princeton University, Princeton, New Jersey  08544 USA}
\author{David A. Huse}
\affiliation{Department of Physics, Princeton University, Princeton, New Jersey  08544 USA}
\author{Waseem S. Bakr}
\email[Corresponding author. Email: ]{wbakr@princeton.edu}
\affiliation{Department of Physics, Princeton University, Princeton, New Jersey  08544 USA}
\date{\today}

\begin{abstract}
Using quantum gas microscopy we study the late-time effective hydrodynamics of an isolated cold-atom Fermi-Hubbard system subject to an external linear potential (a ``tilt").  The tilt is along one of the principal directions of the two-dimensional (2D) square lattice and couples mass transport to local heating through energy conservation. We study transport and thermalization in our system by observing the decay of prepared initial density waves as a function of wavelength $\lambda$ and tilt strength and find that the associated decay time $\tau$ crosses over as the tilt strength is increased from characteristically diffusive to subdiffusive with $\tau\propto\lambda^4$. In order to explain the underlying physics we develop a hydrodynamic model that exhibits this crossover. For strong tilts, the subdiffusive transport rate is set by a thermal diffusivity, which we are thus able to measure as a function of tilt in this regime. We further support our understanding by probing the local inverse temperature of the system at strong tilts, finding good agreement with our theoretical predictions. Finally, we discuss the relation of the strongly tilted limit of our system to recently studied 1D models which may exhibit nonergodic dynamics.
\end{abstract}

\maketitle


\textit{Introduction.}---While non-interacting particles in a tilted lattice potential have been studied for almost a century~\cite{Bloch1929,Wannier1962,Mendez-Hong1988,Voisin-Regreny1988}, the dynamics of strongly tilted and isolated many-body systems with strong interactions have been relatively unexplored. Characterizing the late-time behavior of such closed quantum many-body systems away from equilibrium is a topic of fundamental interest.
In a series of recent papers~\cite{Nahum-Haah2017,Nahum-Haah2018,vonKeyserlingk-Sondhi2018,Khemani-Huse2018,Rakovsky-vonKeyserlingk2018,Gopalakrishnan-Vasseur2018,Banks-Lucas2019} it was shown how irreversible dissipative dynamics can emerge from the unitary evolution of closed quantum systems. Thus generically we expect the transport of conserved quantities in such systems to behave hydrodynamically at late times as long as the system does thermalize. On the experimental front, advances in quantum simulation with cold atoms and other platforms have allowed for unprecedented control of quantum many-body systems, and for the controlled study of their dynamics~\cite{Meinert-Nagerl2014,Choi-Gross2016,Bernien-Lukin2017,Scherg-Aidelsburger2018,Rispoli-Greiner2018,GuardadoSanchez-Bakr2018,Brown-Bakr2019}. For example, in a recent study diffusive charge transport was observed in an isolated strongly-interacting 2D Fermi-Hubbard system~\cite{Brown-Bakr2019}. Here we follow that work by observing the dynamics of the same cold-atom Fermi-Hubbard system subject to a strong external linear potential, or ``tilt", and find a crossover to qualitatively different subdiffusive behavior at strong tilts.

The dynamics of a weakly tilted 2D Fermi-Hubbard model were studied in Ref.~\cite{Mandt-Rosch2011} using semiclassical methods. That work formulated an understanding of the dynamics in which regions with positive local temperature heat up and transport charge ``up" the tilt, and regions with negative local temperature~\cite{Rapp-Rosch2010,Braun-Schneider2013} transport charge ``down" the tilt as the system approaches global equilibrium. Notably, if a tilted quantum system of particles in a lattice does approach thermal equilibrium within a band, that equilibrium is characterized by an infinite temperature~\cite{Mandt-Rosch2011}.  In contrast, recent theoretical works~\cite{vanNieuwenburg-Rafael2019,Schulz-Pollmann2019} explored the prospect of a transition to a localized phase in strongly tilted interacting 1D systems. While some evidence for this was found, it was suggested that this was the result of energetically-imposed local kinetic constraints that conserve the center of mass (COM)---a phenomenon later referred to as ``Hilbert space fragmentation"~\cite{Sala-Pollmann2019,Khemani-Nandkishore2019}. This mechanism for nonergodicity at strong tilts depends on factors such as the range of interactions, the dimensionality of the system, and the direction of the tilt. In what follows, we explore a system which does not exhibit such nonergodicity. Thus this work is most directly related to Refs.~\cite{Brown-Bakr2019,Mandt-Rosch2011}, although initial motivation for this study was derived from Refs.~\cite{vanNieuwenburg-Rafael2019,Schulz-Pollmann2019}, and investigating any nonergodic aspects of tilted systems is an interesting avenue for future work.

In this work we study the effect of an external tilt on the late-time emergent hydrodynamics of a 2D cold-atom system. This is done by varying the tilt strength and observing the relaxation of prepared initial density waves of various wavelengths $\lambda$. We observe a crossover from a diffusive regime at weak tilts, where the relaxation time $\tau$ scales like $\tau \propto \lambda^2$, to a subdiffusive regime at stronger tilts, where $\tau  \propto \lambda^4$. We then construct a hydrodynamic model that exhibits the same crossover and discuss the underlying physics that leads to the subdiffusive transport. Using the hydrodynamic model we extract the tilt-dependent thermal diffusivity of this system. We further verify our understanding of the underlying physics by measuring the local inverse temperature profile of the system, thus confirming a prediction of our theoretical model that this profile should correspond to local equilibrium and be displaced by a quarter wavelength relative to the density profile.

\begin{figure}
\includegraphics[width=\columnwidth]{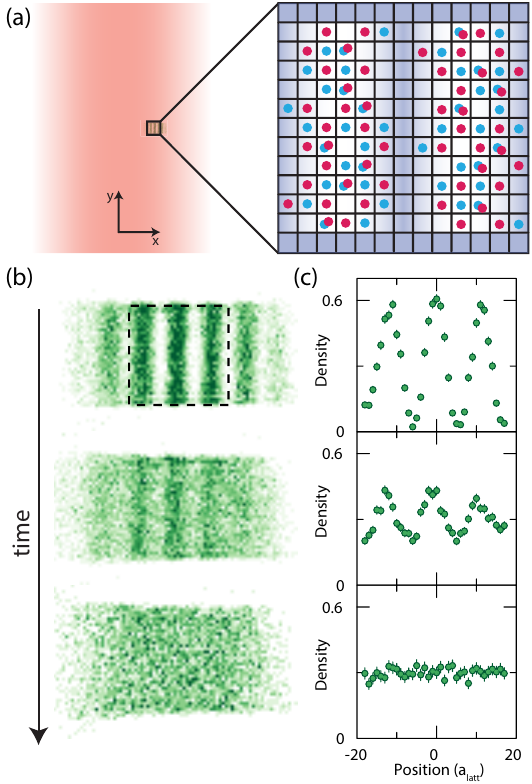}
\caption{{\bfseries Experimental setup and measurements.} (a) An off-centered beam generates a potential at the atoms that is approximately linear in $x$ and independent of $y$. Blue-detuned light projected through a spatial light modulator is used to prepare the initial density waves of our experiments, with tunable wavelength in the direction of the tilt and hard walls a distance of $35\ a_\mathrm{latt}$ apart in the perpendicular direction. The figure is a schematic intended to portray the experimental setup and is not to scale. (b) Density of a single spin component {\it vs.} time averaged over $\sim 10$ images. The dotted square denotes the region of interest (ROI) in which our measurements were taken. (c) Evolution of the {\it y}-averaged density in the ROI of (b) as a function of $x$. The data corresponds to a system with interaction energy $U/t_h = 3.9(1)$, tilt strength $Fa_{\mathrm{latt}}/t_h= 0.99(3)$, and an initial density modulation of wavelength $\lambda/a_{\mathrm{latt}} = 11.46(3)$. The density profile is shown at times $\SI{0}{\milli\second}$ ($0 ~\hbar/t_h$),  $\SI{0.5}{\milli\second}$ ($2.6 ~\hbar/t_h$), and $\SI{15}{\milli\second}$ ($36 ~\hbar/t_h$) from top to bottom. \label{fig:fig1}}
\end{figure}

\textit{System.}---Our system is well-described by the tilted Fermi-Hubbard Hamiltonian $\hat{H} = \hat{H}_\mathrm{FH} - F \hat{N}_f \hat{x}_\mathrm{COM}$ where $\hat{H}_\mathrm{FH}$ is the conventional Fermi-Hubbard Hamiltonian on a square lattice, $F$ is the tilt strength, $\hat{N}_f$ is the total number of fermions, and $\hat{x}_\mathrm{COM}$ is the $x$ component of the COM.  The repulsive on-site interaction energy is denoted by $U$, and the single-particle hopping energy by $t_h$. We emphasize that the system is tilted in only one of the lattice directions, which we denote with $x$.

We realize our tilted 2D Fermi-Hubbard model by loading a balanced mixture of two hyperfine ground states of $^6$Li into an optical lattice~\cite{Brown-Bakr2017}. The tilt is generated by an off-centered $\SI{1064}{\nano\meter}$ Gaussian beam of waist $\sim \SI{180}{\micro \meter}$, as depicted in Fig.~\ref{fig:fig1}\hyperref[fig:fig1]{(a)}. The resulting potential is linear to within $10\%$ across a region of length $40\ a_\mathrm{latt}$ ($\SI{30}{\micro \meter}$), where $a_\mathrm{latt}$ is the spacing of the optical lattice, and the strength of the potential gradient can be tuned from $0$ to $\sim h\times\SI{5.5}{\kilo \hertz} / a_\mathrm{latt}$~\cite{SuppOnline}. The beam is oriented such that the gradient is aligned with one of the two principal axes of the square lattice. A spatial light modulator (SLM) is used to project sinusoidal potentials of tunable wavelength along the direction of the gradient, and also remove any harmonic confinement from trapping potentials in the region of interest, similar to what was done in~\cite{Brown-Bakr2019}. This allows us to prepare initial density modulations of tunable wavelength. We also add ``hard walls" in the direction perpendicular to the gradient in order to contain the atoms in that direction and keep the average density constant over the experimental runtime (see Fig.~\ref{fig:fig1}\hyperref[fig:fig1]{(a)}). 

The atoms are adiabatically loaded into the lattice plus SLM potential at zero gradient (no tilt). The sinusoidal component of the SLM potential is chosen such that the resulting atom-density wave varies spatially with $0.0 \lesssim \avg{\hat{n}_i} \lesssim 1.2$ (see Fig.~\ref{fig:fig1}\hyperref[fig:fig1]{(b-c)}), where $\hat{n}_i=\hat{n}_{i,\uparrow}+\hat{n}_{i,\downarrow}$. We also performed experiments with smaller-amplitude density waves and found no qualitative difference in our results~\cite{SuppOnline}. Once the initial density wave is prepared we suddenly turn off the sinusoidal component of the potential created by the SLM, and turn on the tilt potential, thus initiating the dynamics. We focus on a square region of interest with a size of $35 \times 35$ lattice sites and measure only the single spin component $\avg{\hat{n}_{i,\uparrow}}$  using fluorescence imaging~\cite{Brown-Bakr2017} since in a spin-balanced system $\avg{\hat{n}_i} = 2\avg{\hat{n}_{i,\uparrow}}$.

We performed all experiments at an optical lattice depth of $7.4(1) E_R$, where $E_R / h = \SI{14.66}{\kilo \hertz}$ is the recoil energy and $h$ is Planck's constant. This leads to a hopping rate of $t_h/h = \SI{820(10) }{\hertz}$.  We work at a magnetic field of $\SI{595.29(4)}{\gauss}$ nearby a Feshbach resonance centered on $\SI{690}{\gauss}$. This leads to a scattering length of $472.0(9)\ a_0$, where $a_0$ is a Bohr radius, which translates to an interaction energy of $U/t_h = 3.9(1)$ in the Fermi-Hubbard Hamiltonian. We tune the tilt strength $F$ to values of up to $F a_\mathrm{latt} / t_h \approx 6$ which allows us to explore tilts well above the crossover from diffusive to subdiffusive dynamics.

It is of note that we do not reach tilt strengths so strong that it would be accurate to describe our system over the experimental runtime using an effective Hamiltonian which exactly conserves the COM. Therefore we emphasize that this work does not focus on the physics of fracton-like systems with a strictly conserved dipole moment, nor does it explore the possible nonergodic dynamics in such systems, although these topics are an interesting direction for future research~\cite{vanNieuwenburg-Rafael2019,Schulz-Pollmann2019,Sala-Pollmann2019,Khemani-Nandkishore2019,Pai-Nandkishore2019,Pai-Pretko2019,Nandkishore-Hermele2019}.  However, our tilted system does show an {\it emergent} conservation of the COM in the long wavelength limit where the potential energy of the tilt dominates the conserved total energy.


\textit{Results.}---Our experimental protocol consists of preparing initial density waves of various wavelengths in a potential with tilt $F$ and imaging the system's density profile after it has evolved under its own unitary dynamics for some time $t$.  We analyze our data by averaging all measurements from a certain wavelength, tilt, and time, and we also average the density in the direction perpendicular to the tilt. This yields the averaged density profile along the tilted direction as a function of time, as shown in Fig.~\ref{fig:fig1}\hyperref[fig:fig1]{(c)}.  For each wavelength, tilt, and time we fit the density profile to a sinusoid, $n(x,t)={\bar n} + A(t)\cos{(\phi(t)+2\pi x/\lambda)}$, after adjusting for any small amount of atom loss, with the wavelength being fixed by the fit to the initial profile. We extract both the phase $\phi$ and amplitude $A$ of the sinusoidal fit as a function of time, normalizing the amplitude by its initial value $A(0)$.  The main results of this paper are derived from tracking the decay of the amplitude $A(t)$ with time.

Any change in the phase with time is a result of the distance the center of mass ``falls down" the tilt as the system heats up in the first band of the lattice potential. More precisely, an initial state with energy density corresponding to a finite temperature in the non-tilted Fermi-Hubbard system will evolve down the gradient of the tilted potential. As this happens the tilt does work $\sim F \Delta x_\mathrm{COM}$ per particle for a bulk shift of $\Delta x_\mathrm{COM}$, and this work gets converted locally to kinetic and interaction energy in the system (the $t_h$ and $U$ terms)~\cite{Mandt-Rosch2011}. Since the $t_h$ and $U$ terms can only accommodate up to an energy of order $\sim t_h+U$ per particle before reaching infinite temperature, the shift of the COM of the system cannot be more than $\sim (t_h+U)/F$.  We observe phase changes during the dynamics that are consistent with this approximate bound. We corroborate that the atoms are not excited to higher bands using a technique described in~\cite{Brown-Bakr2019arpes}. 


\begin{figure}
\includegraphics[width=\columnwidth]{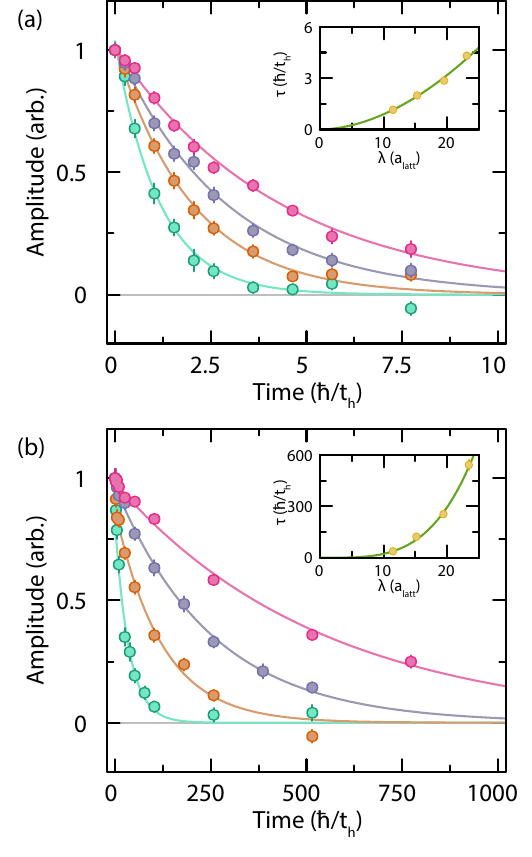}
\caption{{\bfseries Time decay of density waves.} Fitted normalized relative amplitudes of the periodic density modulation (circles) {\it vs.} time for wavelengths $11.46(3)$ (green), $15.16(5)$ (orange), $19.33(7)$ (purple), and $23.3(2)$ (pink) in units of $a_\mathrm{latt}$. The lines are exponential fits to the decay at late times after any initial average heating (phase change). (Insets) Fitted decay times {\it vs.} wavelength (yellow circles) and a power law fit of the form $\tau \propto \lambda^\alpha$ (green line). (a) Dataset for tilt strength $F a_\mathrm{latt} /t_h = 0$. (b) Dataset for tilt strength $F a_\mathrm{latt} /t_h = 2.00(3)$. \label{fig:fig2}}
\end{figure}
 
\begin{figure}
\includegraphics[width=\columnwidth]{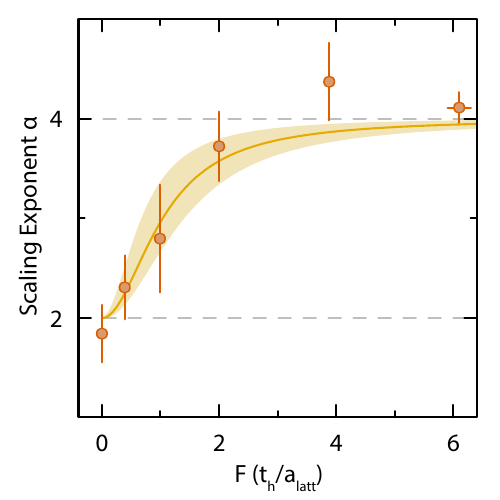}
\caption{{\bfseries Diffusive to subdiffusive crossover.} Extracted scaling exponent $\alpha$ for $\tau \propto \lambda^\alpha$ from datasets at different tilts (orange circles). As the tilt is increased from $F a_\mathrm{latt}/t_h=0$ to $F a_\mathrm{latt}/t_h \approx 6$ the relaxation of initial density waves crosses over from characteristically diffusive ($\alpha=2$) to subdiffusive with $\alpha \approx 4$. The shaded curve is a prediction of our hydrodynamic model which is derived in detail in the Supplement~\cite{SuppOnline}. \label{fig:fig3}}
\end{figure}

At late times we observe an approximately exponential decay of the density modulation (see Fig. \ref{fig:fig2}). We fit an exponential to these curves to extract decay times as a function of $\lambda$ and $F$. This is done at tilts $F a_\mathrm{latt} /t_h \in \{ 0, 0.39(1), 0.99(3), 2.00(3), 3.88(9), 6.1(2) \}$ and for initial density waves with wavelengths $\lambda/a_\mathrm{latt} \in \{ 11.46(3), 15.16(5), 19.33(7), 23.3(2) \}$. We also use $\lambda/a_\mathrm{latt} = 7.69(3)$ for $F a_\mathrm{latt} /t_h \approx 6$ as the decay time of the longest-wavelength modulation becomes very large for this tilt. Decay times that we observe vary increasingly with the tilt strength $F$, from $1\mathrm{-}5\ \hbar/t_h$ at zero gradient up to $10^3\mathrm{-}10^4\ \hbar/t_h$ for $F a_\mathrm{latt} /t_h \approx 6$. At each value of the tilt strength we fit a power law of the form $\tau \propto \lambda^\alpha$ to our measured decay times. Diffusive relaxation has a characteristic $\tau \propto \lambda^2$ dependence ($\alpha =2$), while values of $\alpha > 2$ indicate slower subdiffusive dynamics. Fig. \ref{fig:fig2} shows the full analysis for two of the values of $F$. From the extracted exponents $\alpha$ we observe a crossover from diffusive relaxation at weak tilts, where $\alpha \approx 2 $, to subdiffusive behavior with an exponent of $\alpha \approx 4$ at stronger tilts. This crossover is shown in Fig.~\ref{fig:fig3} along with the theoretical prediction of our hydrodynamic model.

Our observation of diffusive dynamics at weak tilts is consistent with the analysis of Ref.~\cite{Mandt-Rosch2011}, and with the diffusive transport observed in previous experiments on the same system at $F=0$~\cite{Brown-Bakr2019}, albeit at lower temperatures. The crossover to subdiffusion with $\alpha \approx 4$ at strong tilts was, until now, previously unobserved, and its observation and explanation is the main result of this work. Below, and more completely in the Supplement, we construct a hydrodynamic model of our system to help explain these observations. We also further test our understanding of the mechanism behind the subdiffusive transport by experimentally verifying our model's predictions for the local temperature profile.

\textit{Hydrodynamic model.}--- We denote the non-tilt energy density due to $t_h$ and $U$ terms by $e(x,t)$, and the number density of fermions by $n(x,t)$.  Our system is, on average, uniform along the $y$ direction, so
$e$ and $n$ are assumed to only depend on $x$ and $t$.  $n$ is a conserved density and so is $\epsilon=e-Fxn$, the total energy density including the tilted potential. 

For nonzero tilt, our system heats up to near infinite temperature within the lowest band, where the thermodynamic properties are readily calculated using the high-temperature expansion.  There are then three unknown transport coefficients in the most general formulation of our model: diffusivities for each of the two conserved densities and a thermopower coefficient which might be significant for this system since the energy and atom transport are strongly coupled by the tilt.  Our data does not have enough detail to allow us to estimate all three of these transport parameters. However, in the stronger-tilt regime where $\tau \sim \lambda^4$, a tilt-dependent thermal diffusivity is the only transport coefficient that enters in the relaxation, and thus this one parameter can be determined from our measurements. We therefore present a less general version of our model for this strong-tilt regime here, leaving the more general model to the Supplement.

Let us first consider the infinite temperature equilibrium that our system thermalizes to at late times.  This is a limit of zero inverse temperature ($\beta\rightarrow 0$) and infinite chemical potential ($\mu\rightarrow\infty$), with a finite spatially uniform $\beta\mu$; we call this equilibrium value $\bar{\beta\mu}$.  This uniform equilibrium has atom number density $\bar{n}=2e^{\bar{\beta\mu}}/(1+e^{\bar{\beta\mu}})$ per site.  It is convenient when separating the energy into tilt and nontilt terms to choose the interaction term at each site to be $U(n_{\uparrow}-(\bar{n}/2))(n_{\downarrow}-(\bar{n}/2))$.   This choice amounts to changing the total energy and potential $V(x)$ by constants, so it does not change the physics.  With this choice, the equilibrium nontilt energy density vanishes: $\bar{e}=0$.

The density profile at finite time has an additional sinusoidal component: $n(x,t)=\bar{n}+A_0e^{-t/\tau}\cos {kx}$ with $k=2\pi/\lambda$ (choosing the origin so there is no added phase in the argument of the cosine).  In the strong tilt, small $k$ regime we are considering now, this density profile is at {\it local} equilibrium with a time-dependent and spatially nonuniform inverse temperature $\beta(x,t)$.  We assume the system is also near global equilibrium, so we work to lowest order in $A_0$ and $\beta$.  Near position $x$, if we have local equilibrium in the tilted potential $V(x)=-Fx$ in this high temperature limit, the density is given by $n(x)=2e^{\beta(\mu+Fx)}/(1+e^{\beta(\mu+Fx)})$. So, in the long wavelength limit we are considering here, the density gradient is $dn/dx=Fn(1-(n/2))\beta(x)$.  Thus to leading order the temperature profile is given by $-A_0ke^{-t/\tau}\sin{kx}=F\bar{n}(1-(\bar{n}/2))\beta(x,t)$. Using this result along with a high temperature expansion to write $e$ as a function of $\beta$ to leading order, we obtain the nontilt energy profile
\begin{equation}
e(x,t)=\frac{A_0}{F}\left( 4t_h^2+U^2\frac{\bar{n}}{4} \left( 1-\frac{\bar{n}}{2} \right) \right) k e^{-t/\tau}\sin{kx}
\end{equation}
at local equilibrium to lowest order in $A_0$ and $k$. Now that we have determined the profiles of $n$ and $e$ assuming local equilibrium, next we consider the dynamics and use energy and number conservation to determine the relaxation time $\tau$. In the regime we are now considering, the rate-limiting bottleneck is the transport of nontilt energy (heat) through the system. This limits the rate at which tilt energy can be converted to heat and dissipated to the rest of the system, and thus the rate at which the whole system relaxes.

The relaxation of the number density implies, via the continuity equation for atom number, an atom number current density of 
\begin{equation}
\label{eqn:number_current}
j_n(x,t)=\frac{A_0}{k\tau}e^{-t/\tau}\sin{kx}~.
\end{equation}
This current flows along the tilt direction, locally converting tilt energy to nontilt energy.  In addition, there is a heat current $j_h(x,t) = - D_\mathrm{th} \nabla e(x,t)$ flowing due to the temperature gradients, where $D_\mathrm{th}(F)$ is a tilt-dependent thermal diffusivity. Conservation of energy is then
\begin{equation}
\label{eqn:energy_cons}
\dot{e}=D_{th}\nabla^2 e + Fj_n~,
\end{equation}
showing the contribution of heat diffusion and the conversion of energy from tilt to nontilt due to the atom current $j_n$. In the strong tilt regime we are considering, the two terms on the RHS of Eqn.~\ref{eqn:energy_cons} are each much larger in magnitude than the LHS: the motion of the atoms converts tilt energy to nontilt energy and this is dissipated by thermal transport, while the amplitude of the inhomogeneities decays slowly ($D_\mathrm{th} k^2 \tau \gg 1$).  In this strong tilt regime, the decay rate is 
\begin{eqnarray}
\label{eqn:tau}
\frac{1}{\tau}=\frac{D_\mathrm{th} k^4}{F^2} \left( 4t_h^2+U^2\frac{\bar{n}}{4} \left( 1-\frac{\bar{n}}{2} \right) \right) \ll D_\mathrm{th}k^2~,
\end{eqnarray}
and the condition for the validity of this regime is
\begin{eqnarray}
\label{eqn:validity}
k^2 \left( 4t_h^2+U^2\frac{\bar{n}}{4} \left( 1-\frac{\bar{n}}{2} \right) \right)\ll F^2~.
\end{eqnarray}

\begin{figure}
\includegraphics[width=\columnwidth]{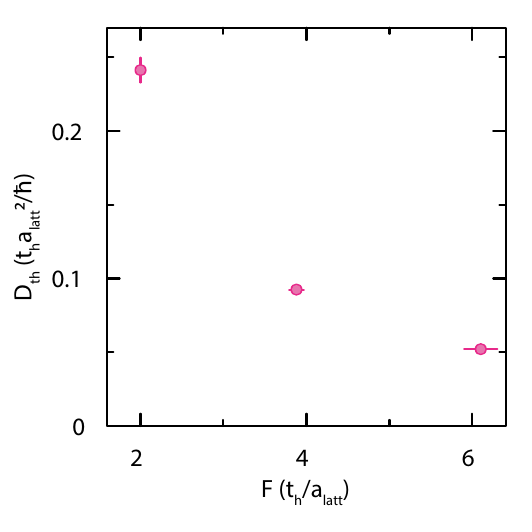}
\caption{{\bfseries Thermal diffusivity.} Extracted thermal diffusivity (circles) {\it vs.} gradient. The values were extracted by doing a fit of our hydrodynamic model to all wavelengths of each gradient simultaneously~\cite{SuppOnline}. \label{fig:fig4}}
\end{figure}

We use Eqn.~\ref{eqn:tau} to extract the thermal diffusivity $D_\mathrm{th}$ as a function of tilt strength in the regime consistent with $\tau \propto \lambda^4$ and plot the result in Fig.~\ref{fig:fig4}.
From the validity condition of Eqn.~\ref{eqn:validity} we can also estimate the location of the crossover shown in Fig.~\ref{fig:fig3}. Plugging in the experimental values of $U/t_h=4$ and $\bar{n}=0.6$, and any value of $k$ from the experimental range $k a_\mathrm{latt} \in [2\pi / 24, 2\pi / 12]$, we get the condition that $\alpha \approx 4$ when $F a_\mathrm{latt} / t_h \gg 1$, which is consistent with the data shown in Fig.~\ref{fig:fig3}. A more complete model is detailed in the Supplement, and this model is used to derive the superimposed curve of Fig.~\ref{fig:fig3} which agrees quantitatively with our experimental results.  This more detailed model also gives the thermal diffusivity $D_\mathrm{th}$ in terms of all of the transport coefficients, including the thermopower.  We therefore conclude that our hydrodynamic model captures the essential physics leading to the main observation of this paper: the crossover from diffusive to subdiffusive relaxation with $\tau \propto \lambda^4$ as the tilt becomes strong.

\begin{figure*}
\includegraphics[width=2\columnwidth]{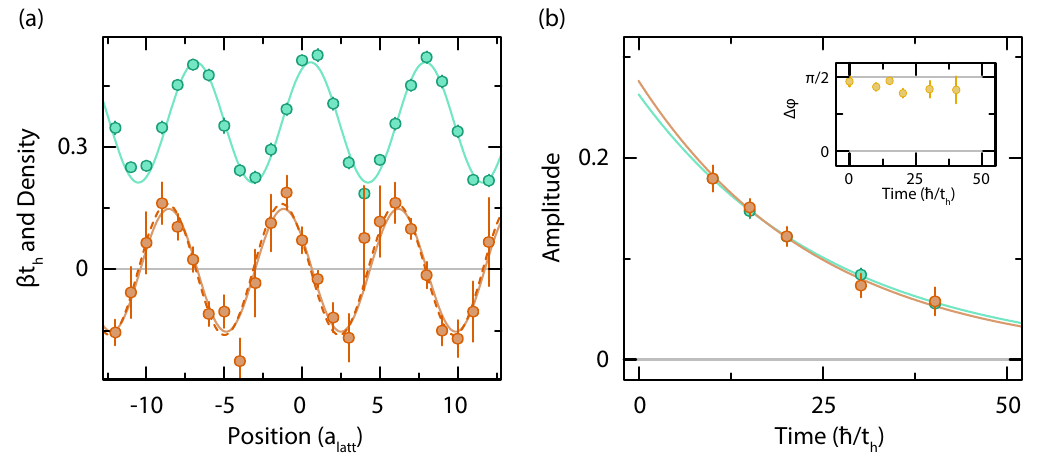}
\caption{{\bfseries Local inverse temperature.}  Near infinite temperature, the density of singles can be used for thermometry. For a tilt strength of $Fa_{\mathrm{latt}}/t_h = 3.4(1)$ and periodic modulation of wavelength $7.69(3)\ a_\mathrm{latt}$, we measure the average single component density (green) and the density of singles (not shown) in order to extract the local inverse temperature of the cloud (orange). (a) The measured average single component density (green circles) and extracted inverse temperature $\beta t_h$ (orange circles) with sinusoid fits (solid lines) after a decay time of $15.1\ \hbar/t_h$. In the case of the inverse temperature, the dashed line is the predicted inverse temperature profile from the density fit and local equilibrium. (b) The amplitude of the density (green) and inverse temperature (orange) modulations {\it vs.} time (circles) with exponential decay fits (solid lines). (inset) Shows the phase difference of the sinusoid fits between the single component density and the extracted local inverse temperature {\it vs.} time (yellow circles). \label{fig:fig5}}
\end{figure*}

The picture we have laid out in this section is one where, at strong tilts and long wavelengths, the system quickly achieves local equilibrium, locking the local inverse temperature to the density profile. As the density profile decays, local number density currents flow, and by conservation of energy this necessitates the flow of nontilt energy in the system. It is this flow of nontilt energy that we claim bottlenecks the relaxation in the large $F$ regime, and thus $D_\mathrm{th}$ sets the relaxation rate of the system.  A prediction of this understanding is {\it local} equilibrium between $\beta(x,t)$ and $n(x,t)$.  We verify this prediction by measuring the single component density and singlon occupancy profiles in our system and solving for the inverse temperature in the atomic limit, which is an effective method of thermometry at such high temperatures. In Fig.~\ref{fig:fig5}\hyperref[fig:fig5]{(a)} and \hyperref[fig:fig5]{(b)} we show both the density and local inverse temperature profiles, the decay of both of their amplitudes, and the phase difference between them in time (inset). From this we see that the $\beta$ profile is at local equilibrium, locked at a quarter wavelength phase shift from the density profile, and both profiles decay together in time, as predicted by our understanding of the subdiffusive regime of this system.

\textit{Summary and outlook.}---We studied a new regime of thermalization in a square-lattice cold-atom Fermi-Hubbard system subject to an external linear potential.  Our system was effectively closed and evolved under its own unitary dynamics starting from prepared initial density waves of various wavelengths $\lambda$.  By observing how the amplitude of these initial density modulations evolved in time we found two qualitatively different hydrodynamic regimes and a crossover between them:  At weak tilts the system relaxes diffusively, in accordance with previous theory~\cite{Mandt-Rosch2011} and experiments~\cite{Brown-Bakr2019}.  At strong tilts, we found a new regime where the system relaxes subdiffusively with a decay time $\tau$ that scales as $\tau \propto \lambda^4$.  We argued that this subdiffusive behavior is a result of having to ``drain" the large reservoir of tilt energy via the bottleneck of heat transport en route to {\it global} equilibrium, and is captured effectively by a hydrodynamic description with the system remaining near {\it local} equilibrium.  To test this understanding we measured the local temperature profile and do indeed find that the system remains near local equilibrium as it relaxes in this subdiffusive regime.  In the Supplement, we also develop and present a more complete and detailed hydrodynamic model that quantitatively captures the crossover between the diffusive and subdiffusive regimes (Fig.~\ref{fig:fig3}). In the strongly tilted regime we used our model to extract the tilt-strength-dependent thermal diffusivity that bottlenecks the relaxation of the system. One perspective on why this novel subdiffusive regime appears is that in the strong-tilt and long-wavelength limit the center-of-mass potential energy is the dominant part of the total energy, so energy conservation becomes an emergent almost-conservation of the center of mass.

In contrast to recent theoretical studies of potential ergodicity breaking in tilted 1D systems~\cite{vanNieuwenburg-Rafael2019,Schulz-Pollmann2019}, in this work we focused on the novel effects of a tilt on the approach to equilibrium in an isolated system that does indeed thermalize.  This thermalization was robust because our system had a tilt potential along only one of the two principal axes of the lattice, and the resulting unconstrained motion of atoms in the perpendicular direction produced good thermal baths in each such row of the lattice.  To arrest this thermalization more microscopically, one avenue of future exploration will be to apply tilt potentials along both axes of the lattice to suppress such local thermalization.

\textit{Acknowledgments.}---We thank Vedika Khemani for helpful discussions. This work was supported by the NSF (grant no. DMR-1607277), the David and Lucile Packard Foundation (grant no. 2016-65128), and the AFOSR Young Investigator Research Program (grant no. FA9550-16-1-0269). W.S.B. was supported by an Alfred P. Sloan Foundation fellowship. A.M. acknowledges the support of the Natural Sciences and Engineering Research Council of Canada (NSERC).  D.A.H. was supported in part by the DARPA DRINQS program.



%


\appendix
\pagebreak
\clearpage
\setcounter{equation}{0}
\setcounter{figure}{0}

\renewcommand{\theparagraph}{\bf}
\renewcommand{\thefigure}{S\arabic{figure}}
\renewcommand{\theequation}{S\arabic{equation}}

\onecolumngrid
\flushbottom

\pagebreak
\section*{Supplemental material}

The experimental setup and basic parameters are already described in detail in the supplement of Ref.~\cite{Brown-Bakr2017}. The spatial light modulator calibration is also explained in the supplement of Ref~\cite{Brown-Bakr2019}.

\subsection*{Tilt potential calibration}
To calibrate the gradient and characterize its homogeneity across the region of interest, we used the SLM to prepare an initial state consisting of three thin stripes  of width $\sim 1\ a_\mathrm{latt}$ and a separation of $\sim 20\ a_\mathrm{latt}$, with their long direction oriented orthogonal to the tilt direction. Each stripe consists of a spin-polarized gas of the lowest hyperfine ground state of $^6$Li.

\begin{figure}[h]
\includegraphics[width=\columnwidth]{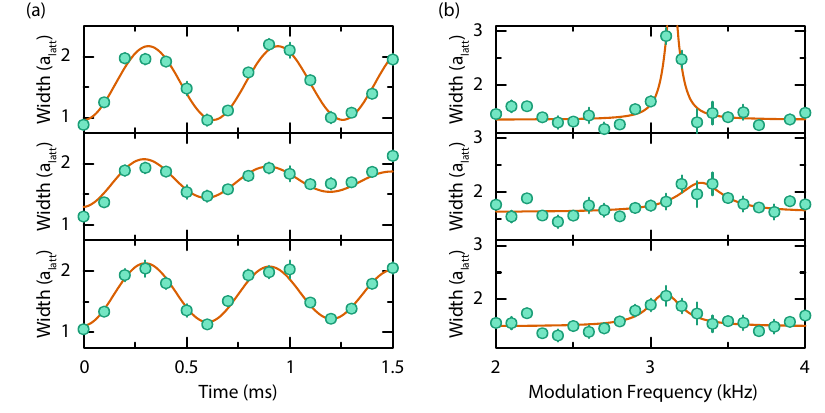}
\caption{{\bfseries Tilt potential calibration.} (a) Bloch oscillation method for characterization of tilt strengths. Each graph corresponds to a measurement of the local gradient at the position of one of the three stripes. The measured tilt strength is $F  a_{\mathrm{latt}} = h\times\SI{1.64(3)}{\kilo \hertz}$ with a maximal difference of $4.6\%$ between stripes. (b) Lattice modulation method for characterization of tilt strengths. The measured tilt strength is $F a_{\mathrm{latt}} = h\times\SI{3.19(7)}{\kilo \hertz}$ with a maximal difference of $7.5\%$ between stripes. \label{fig:sfig1}}
\end{figure}

For weak tilts, we are able to directly measure Bloch oscillations of these non-interacting particles. We do so by fitting a Gaussian profile to the density profile integrated along the direction perpendicular to the tilt which is used to quantify the ``breathing'' oscillation of the width of the stripes. This is similar to what was done in \cite{Preiss-Greiner2015}. From the theory of Bloch oscillations, we expect the width of each stripe to oscillate with a maximal half-width of $A=4t_h/F$ and a period of $T = h/F a_{\mathrm{latt}}$. Thus, by fitting a sinusoid to the evolution of the width of each stripe, we can extract the tilt strength at their respective positions. Fig.~\ref{fig:sfig1}(a) shows an example of such oscillations.

For stronger tilts, directly measuring the Bloch oscillations becomes challenging due to their small amplitude.  Instead we use a modulation technique analogous to what was done in~\cite{Ma-Greiner2011}. We modulate the lattice potential at frequencies on the order of the tilt strength. This brings lattice sites that were decoupled due to the tilt into resonance which results in photon-assisted tunneling. We again measure the width of the thin stripes versus modulation frequency and observe a broadening of the stripes at resonance. Fig.~\ref{fig:sfig1}\hyperref[fig:sfig1]{(b)} shows an example of such a measurement.

We corroborated that for the same potential strength at intermediate tilts, the gradient extracted using the two techniques agrees.

\subsection*{Linear Response}
Our hydrodynamic model assumes linearity in the amplitude of the initial inhomogeneities. In this experiment, we worked with relatively large amplitude density modulations. In a previous study (\cite{Brown-Bakr2019}), we worked with very small amplitude modulations and fit to a linear hydrodynamic model we developed. In the ``tilted'' system studied in this work, we are no longer working close to a ground state, and as such, the strength of the modulation is not expected to be as important.

\begin{figure}[h]
\includegraphics[width=\columnwidth]{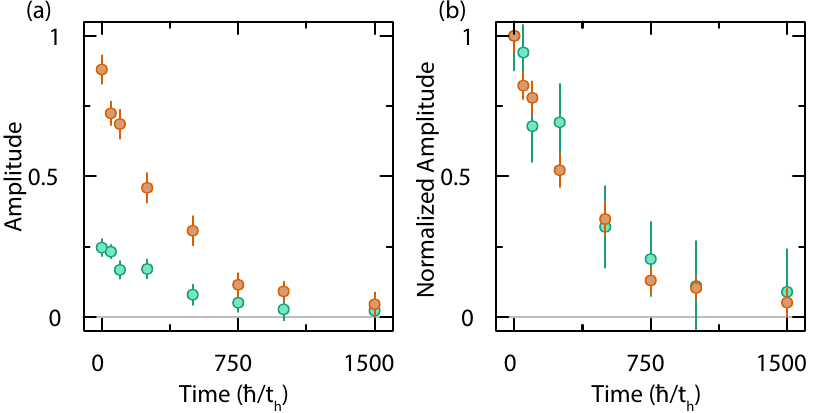}
\caption{{\bfseries Test of linear response.} Decay of the amplitude of the density modulation {\it{vs.}} time for two different initial amplitudes of the modulation. Here $\lambda=11.46(3) a_{\mathrm{latt}}$, $Fa_{\mathrm{latt}}/t_h = 6.1(2)$ and $U/t_h=3.9(1)$. (a) Shows the amplitudes. (b) Shows the amplitudes normalized to the baseline at $t=0$.\label{fig:sfig2}}
\end{figure}

Fig.~\ref{fig:sfig2} shows a comparison between the decay of strong and weak density modulations in a tilted potential. We observe that when we normalize the sinusoid amplitude and look at its decay, there is no measurable difference between the decays within the errorbars. This justifies working with strong modulations in this work to reduce the statistical error in the measurements for a fixed number of repetitions.

\subsection*{Complete hydrodynamic model}
We write our hydrodynamic theory in terms of the particle number density $n(x,t)$ and nontilt energy density $e(x,t)$, as well as their corresponding currents $j_n(x,t)$ and $j_e(x,t)$. Total particle number and total energy are conserved, and these conservation laws can be written as
\begin{eqnarray}
\dot{n}+\nabla \cdot j_n=0 \label{eqn:n_cons}\\
\dot{e} +\nabla \cdot j_e  - F j_n = 0. \label{eqn:e_cons}
\end{eqnarray}

The entropic ``force" laws that describe how currents are driven in this system are of the form $j_e = M_e \chi_e + M_{ne} \chi_n$ and $j_n = M_n \chi_n + M_{en} \chi_e$, where $\chi_e$ and $\chi_n$ are the entropic forces determined by the profiles of $e$ and $n$, and we insist on writing the forces in a ``canonical basis" for which Onsager's reciprocal relations take the simple form $M_{en}=M_{ne}$. The $M$ coefficients are dynamical coefficients that are, in general, difficult to determine from the microscopic model. The off-diagonal coefficient $M_{ne}$ is associated with thermopower-type effects in our system, thus this model is quite general aside from its assumption of linearity, which is well-supported by our experimental measurements. In this canonical basis, the force $\chi_e$ is determined by the local change of entropy when an infinitesimal current of nontilt energy flows but no particle current flows. The force $\chi_n$ is determined in a similar fashion, with an infinitesimal particle current and no nontilt energy current, but note that if an infinitesimal particle current flows, then due to energy conservation there must be a production (or depletion) of nontilt energy. Thus the forces in this system take the form
\begin{eqnarray}
\chi_e &=& \nabla \left( \frac{\partial s}{\partial e} \right)\\
&=& -s_{ee} \nabla e + s_{ne} \nabla n\\
\chi_n &=& \nabla \left( \frac{\partial s}{\partial n} \right) + F \left( \frac{\partial s}{\partial e} \right)\\
&=& -s_{nn} \nabla n + s_{ne} \nabla e -  s_{ee} F (e - \bar{e}(\bar{n})) + s_{ne} F(n-\bar{n}),
\end{eqnarray}
where $s$ is the entropy density of the Fermi-Hubbard model, and we have expanded this entropy density near infinite-temperature equilibrium with $n=\bar{n}$ and $e=\bar{e}(\bar{n})$. The coefficients $s_{ee}$, $s_{nn}$, and $s_{ne}$ come from this high-temperature expansion:
\begin{eqnarray}
s \approx s(\bar{n},\bar{e}(\bar{n})) + s_n (n-\bar{n}) - \frac{1}{2} s_{nn} (n-\bar{n})^2 + s_{ne} (n-\bar{n})(e-\bar{e}(\bar{n})) - \frac{1}{2} s_{ee} (e-\bar{e}(\bar{n}))^2,
\end{eqnarray}
and we emphasize that these coefficients are known functions of $U$, $t_h$ and $\bar{n}$ for the Fermi-Hubbard model. 

Now that we have specified our model, we proceed in determining its eigenmodes and respective relaxation rates, with a particular focus on the slowest mode, which is representative of the late-time behavior that we analyze in the main text. To do this, we first organize our model into a matrix eigenvalue problem. The eigenmodes are functions of definite wavelength $\lambda$, and they decay to equilibrium at a rate $\tau^{-1}$, i.e. $e(x,t)-\bar{e}(\bar{n}) = \mathrm{e}^{- \Gamma t } (a \cos kx + b \sin kx)$ and $n(x,t) - \bar{n} = \mathrm{e}^{- \Gamma t} (c \cos kx + d \sin kx)$, where $k=2\pi /\lambda$ and $\Gamma = 1/\tau$. We therefore write the above deviations of $e$ and $n$ from global equilibrium as $\vec{\rho} = \left( a~b~c~d \right)^T$ in the basis $\{ \mathrm{e}^{- \Gamma t } \cos kx, \mathrm{e}^{- \Gamma t} \sin kx, \mathrm{e}^{-\Gamma  t} \cos kx, \mathrm{e}^{-\Gamma  t } \sin kx\}$. In this language the currents are driven according to $\vec{j}=M U \vec{\rho}$ and the conservation laws are written as $-\Gamma \vec{\rho} = -\tilde{\nabla} \vec{j}$, where
\begin{eqnarray}
U = 
\begin{pmatrix}
0 & -s_{ee} k & 0 & s_{ne} k \\
s_{ee} k & 0 & -s_{ne} k & 0 \\
-s_{ee} F & s_{ne} k & s_{ne} F & -s_{nn} k \\
-s_{ne} k & -s_{ee} F & s_{nn} k & s_{ne}F
\end{pmatrix},~
M = 
\begin{pmatrix}
M_e & 0 & M_{ne} & 0 \\
0 & M_e & 0 & M_{ne} \\
M_{ne} & 0 & M_n & 0 \\
0 & M_{ne} & 0 & M_n
\end{pmatrix},~
\tilde{\nabla} = 
\begin{pmatrix}
0 & k & -F & 0 \\
-k & 0 & 0 & -F \\
0 & 0 & 0 & k \\
0 & 0 & -k & 0
\end{pmatrix}.
\end{eqnarray}
Thus our model is solved via the eigenvalue problem $\Gamma \vec{\rho}_\Gamma = \tilde{\nabla} M U \vec{\rho}_\Gamma$. There are two solutions for $\Gamma$, each with a multiplicity of two corresponding to pure $\cos$ and $\sin$ waves for $n(x,t)-\bar{n}$. The only solution we need to consider at late times is the slow mode with $\Gamma = \Gamma_-$ and $n(x,t) - \bar{n} \propto \cos kx$. This eigenmode is representative of the dynamics of all monochromatic initial conditions at late times. In what follows we discuss some important features of the slowest eigenmode. 

In the limit of small $F$ (and/or large $k$) the slowest mode is diffusive, i.e. $\Gamma \propto k^2$. In the limit of large $F$ (and/or small $k$) the slowest decay rate is
\begin{equation}
\Gamma_- \approx \frac{D_\mathrm{th}}{F^2} \left( \frac{s_{nn}}{s_{ee}} - \frac{s_{ne}^2}{s_{ee}^2} \right) k^4, \label{eqn:gammaminus}
\end{equation}
where $D_\mathrm{th} = \left( M_e - ( M_{ne}^2 / M_n ) \right) s_{ee}$ is the thermal diffusivity, and we will discuss why we identify it as such below. Thus we see that our model crosses over from diffusive to subdiffusive with $\tau \propto \lambda^4$ as $1/F\lambda$ becomes small.

If we assume a scaling of the form $\Gamma_- \propto k^\alpha$ we can estimate the exponent $\alpha$ by $\alpha = \left. \frac{d \log \Gamma_-}{d \log k} \right|_{k=k_e}$ evaluated at some $k$ in the experimental range $k \in [2\pi / 24, 2\pi / 12]$ denoted $k_e$. The general expression for $\alpha$ evaluated this way depends on the dynamical coefficients $M$, but in the limit where $\frac{M_e}{M_n} , \frac{M_{ne}}{M_n} \ll \frac{s_{nn}}{s_{ee}}$ this dependence drops out and we get a parameter-free estimate of $\alpha$ as a function of $F$. In this limit
\begin{equation}
\label{eqn:alpha}
\alpha(F) = 2+\frac{2}{1 + \frac{s_{nn}}{s_{ee}} \frac{k_e^2}{F^2} },
\end{equation}
and this is the theoretical estimate of $\alpha(F)$ that we use to compare to experimental results in the main text (Fig.~\ref{fig:fig3}).

Now we examine the structure of the slowest eigenmode itself and explain why we identify $D_\mathrm{th}$ as mentioned above. At small $k/F$, to leading order, the slowest eigenmode has
\begin{equation}
\label{eqn:slowmode}
\vec{\rho}_{\Gamma_-} = 
\begin{pmatrix}
\frac{s_{ne}}{s_{ee}}\\
\left(\frac{s_{nn}}{s_{ee}} - \frac{s_{ne}^2}{s_{ee}^2} \right) \frac{k}{F}\\
1\\
0
\end{pmatrix},~
\vec{j}_{\Gamma_-}=
\begin{pmatrix}
- \left( M_e - \frac{M_{ne}^2}{M_n} \right) s_{ee} \left( \frac{s_{nn}}{s_{ee}} - \frac{s_{ne}^2}{s_{ee}^2} \right) \frac{k^2}{F}\\
 \left( M_e - \frac{M_{ne}^2}{M_n} \right) s_{ne} \left( \frac{s_{nn}}{s_{ee}} - \frac{s_{ne}^2}{s_{ee}^2} \right) \frac{k^3}{F^2}\\
 0\\
 -\left( M_e - \frac{M_{ne}^2}{M_n} \right) s_{ee} \left( \frac{s_{nn}}{s_{ee}} - \frac{s_{ne}^2}{s_{ee}^2} \right) \frac{k^3}{F^2}
\end{pmatrix}.
\end{equation}
We see that in this mode a modulation of number density with amplitude $\mathcal{O}(1)$ comes with a slow subdiffusive number density current $j_n \propto k^3$ that is ``out of phase" by a quarter wavelength. This number current converts tilt energy to nontilt energy and this generates a small out of phase, nontilt energy profile with amplitude $\propto k/F$. That nontilt energy diffuses and we see that the ratio of amplitudes of the resulting ``in phase" (with $n(x,t)$) energy current to the energy profile it is depleting is $|j_e| / |e| = D_\mathrm{th} k$ with $D_\mathrm{th} = \left( M_e - ( M_{ne}^2 / M_n ) \right) s_{ee}$ as mentioned earlier. This is why we identify $D_\mathrm{th}$ as such. The process of diffusing the nontilt energy that is generated by the particle current that is relaxing the density profile is the bottleneck process and obeys a diffusion equation with diffusivity $D_\mathrm{th}$. That is why this diffusivity shows up as the one unknown coefficient in $\Gamma_-$, and thus we use our data to determine it in the regime where $\tau \propto \lambda^4$ where this mechanism is valid.

Now let's address the $\beta$ profile in this mode. The ``in phase" component of $e-\bar{e}(\bar{n})$ shown in Eqn.~\ref{eqn:slowmode}, which is larger than the out of phase component by a factor of $F/k$, is due to the difference between $\bar{e}(\bar{n})$ and $\bar{e}(n)$, and not due to a nonzero $\beta$ component that is in phase. Since $\beta$ is proportional to $e-\bar{e}(n)$ at high temperatures, to leading order in the high-temperature limit $\beta(x,t)$ is set by the out of phase component of $e-\bar{e}(n)$ which is the same as $e-\bar{e}(\bar{n})$ for that component because the out of phase component of $n-\bar{n}$ is zero by definition (since ``in phase" and ``out of phase" are defined relative to the $n(x,t)$ profile here). Thus this model predicts an out of phase local $\beta$ modulation with amplitude
\begin{eqnarray}
\mathrm{amp}(\beta(x,t)) &=& -s_{ee} \left( \frac{s_{nn}}{s_{ee} - \frac{s_{ne}^2}{s_{ee}^2}} \right) \frac{k}{F}\\
&=& \frac{1}{\bar{n} \left( 1-\frac{\bar{n}}{2} \right) } \frac{k}{F},
\end{eqnarray} 
where we have used the high temperature expressions for $s_{ee}$, $s_{ne}$, and $s_{nn}$. Indeed in the main text we show measurements of the local $\beta$ that are consistent with this prediction (Fig.~\ref{fig:fig5}). We call the nontilt energy current that results from this local $\beta$ profile the ``heat current" $j_h$. Thus $D_\mathrm{th}$ is the diffusivity corresponding to the heat current that is being driven by the nontilt energy that is generated by the relaxation of the particle number distribution.

\subsection*{High temperature expansion}
We compute the grand partition function of the Fermi-Hubbard model in the high temperature expansion to second order in $\beta$ and evaluate the second partial derivatives of the entropy density with respect to $n$, the particle number density, and $e$, the energy density due to $t_h$ and $U$ terms, in order to compute the coefficients $s_{ee}$, $s_{nn}$, and $s_{ne}$. The results are
\begin{eqnarray}
s_{ee} = \frac{16}{\bar{n} (2-\bar{n}) (32 t_h^2 + \bar{n} (2-\bar{n}) U^2)}\\
s_{nn} = \frac{64 t_h^2 + 2 \bar{n} (2+\bar{n}) U^2}{\bar{n} (2-\bar{n}) (32 t_h^2 + \bar{n} (2-\bar{n}) U^2)}\\
s_{ne} = \frac{8 \bar{n} U}{\bar{n} (2-\bar{n}) (32 t_h^2 + \bar{n} (2-\bar{n}) U^2)}.
\end{eqnarray}

For the experimental parameters $\bar{n} = 0.6$ and $U/t_h = 3.9$ these coefficients take the values $s_{nn} \approx 2.96$, $s_{ee} \approx 0.43$, $s_{ne} \approx 0.50$ in units where $t_h=1$. 

\subsection*{Simultaneous fitting of model}

As explained in the previous sections, there is a fast and a slow exponential decay solution to our hydrodynamic model. In the strong tilt regime, Eqn.~\ref{eqn:gammaminus} shows that the slow decay depends only on the thermal diffusivity $D_\mathrm{th}$.

\begin{figure}[h]
\includegraphics[width=\textwidth]{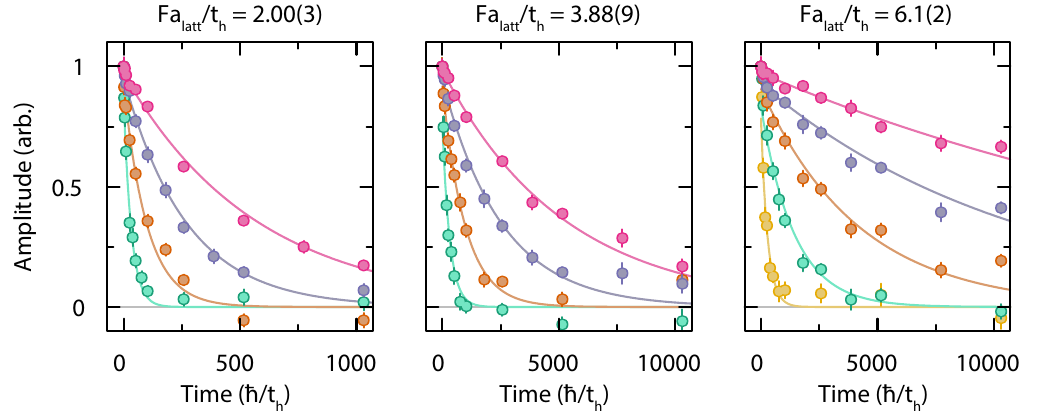}
\caption{{\bfseries Simultaneous fitting of hydrodynamic model.} Fitted normalized relative amplitudes of the periodic density modulation (circles) {\it vs.} time for wavelengths $7.69(3)\ a_\mathrm{latt}$ (yellow), $11.46(3)\ a_\mathrm{latt}$ (green), $15.16(5)\ a_\mathrm{latt}$ (orange), $19.33(7)\ a_\mathrm{latt}$ (purple), and $23.3(2)\ a_\mathrm{latt}$ (pink) at different tilts. The lines are simultaneous fits of the hydrodynamic model to the long-time decay after the initial average heating (phase change). We are able to extract the thermal diffusivity through this fitting method. \label{fig:sfig3}}
\end{figure}

We perform a simultaneous fit to all wavelengths at a given tilt strength as explained in the supplement of \cite{Brown-Bakr2019}. The fitting function is
\begin{eqnarray}
    A(t) &=& A_0 e^{-\Gamma_{-}(D_\mathrm{th}, F, k) t},
\end{eqnarray}
and it is fitted only to the late-time decay. Here, $A_0$ is a fitting parameter that can vary for each wavelength while $D_\mathrm{th}$ is fitted globally to all wavelengths. The parameters $F$ and $k$ are fixed according to our experimentally measured values. The results of fitting this model to measurements in the strong tilt regime are shown in Fig.~\ref{fig:sfig3}.

\end{document}